# A Regime Shift in Atlantic Surface Currents Reveals a Step-like Decline of the Meridional Overturning Circulation

Han Huang[1], Ningning Tao[1], Hongyu Wang[1], Teng Liu[1,4,5], Fei Xie[1], Xichen Li[7], Yongwen Zhang[3], Niklas Boers[4,5], Jingfang Fan[1,5], Deliang Chen[6], Xiaosong Chen[1,2*]

[1] *School of Systems Science & Institute of Nonequilibrium Systems, Beijing Normal University, Beijing, China*

[2] *Institute for Advanced Study in Physics and School of Physics, Zhejiang University, Hangzhou, China*

[3] *Data Science Research Center, Faculty of Science, Kunming University of Science and Technology, Kunming, Yunnan, China*

[4] *Earth System Modeling, School of Engineering and Design, Technical University of Munich, Ottobrunn, Germany*

[5] *Potsdam Institute for Climate Impact Research, Potsdam, Germany*

[6] *Department of Earth System Science, Tsinghua University, Beijing, China*

[7] *Laboratory for Climate and Ocean-Atmosphere Studies, Department of Atmospheric and Oceanic Sciences, Peking University, Beijing, China*

\* Corresponding authors:

Prof. Xiaosong Chen (chenxs@zju.edu.cn)

**Abstract:**

The Atlantic surface currents associated with the Atlantic Meridional Overturning Circulation (AMOC) play a central role in regulating Earth's climate, yet their large-scale dynamical response to climate variability remains poorly understood. Here we identify a previously unrecognized basin-scale phase of Atlantic surface circulation, termed the Atlantic Convergence–Divergence Mode (ACDM), characterized by a convergence–divergence pattern in the North Atlantic and coherent meridional flows in the South Atlantic. We show that the ACDM experienced a pronounced regime shift in 2009, marked by weakened vertical water exchange and reduced meridional transport. This transition closely coincides with direct RAPID–MOCHA AMOC observations and is driven by AMOC-modulated multi-scale forcing: a low-frequency oceanic thermal reorganization that preconditions the system, and episodic atmospheric shocks that trigger the shift. By identifying the ACDM variability as a sensitive and physically grounded proxy for interannual AMOC fluctuations, we reveal that the observed 2009 shift signifies a nonlinear, step-like weakening of AMOC that triggered a fundamental basin-scale reorganization of Atlantic surface currents. Our results offer a dynamical explanation for the AMOC's recent decline and demonstrate its inherently non-linear nature, highlighting the need to account for step-like transitions in assessing its stability and future evolution.

**Significance:**

The Atlantic Meridional Overturning Circulation (AMOC) is a critical tipping element in the Earth's climate system whose collapse would trigger irreversible global catastrophes. Currently, a critical discrepancy exists: traditional proxies fail to align with direct observations on interannual scales, leaving the AMOC's actual stability highly uncertain, especially for the potential of abrupt changes. We resolve this by identifying a basin-scale surface response with a complexity-science framework, the Atlantic Convergence-Divergence Mode (ACDM), which provides precise proxy for interannual AMOC variability. Our findings reveal that instead of a gradual decline, the AMOC underwent a sudden, step-like decline around 2009 and further triggers a reorganization in Atlantic surface currents, highlighting the urgent need to account for abrupt, step-like transitions in assessing Earth system stability.

**Main**

**Introduction**

The global ocean circulation is a fundamental component of the Earth's climate system, governing the redistribution of heat, energy, and carbon across the planet (1–5). Among the major ocean basins, the Atlantic plays a uniquely prominent role owing to its distinctive topography, which connects the North and South Poles, and its meridional overturning circulation, commonly referred to as the Atlantic Meridional Overturning Circulation (AMOC) (3, 6–8). The AMOC is widely regarded as one of the most important processes in the Earth system because of its critical influence on global climate and, in particular, meridional heat transport (9–19). Paleoclimate evidence indicates that abrupt climate events, such as Dansgaard–Oeschger oscillations and associated rapid temperature shifts, were closely linked to AMOC variability (14, 20). Conceptual and numerical models further demonstrate that the AMOC can exhibit bistability, with the positive salt-advection feedback identified as a key underlying mechanism (13–15, 21). At present, both models and observations suggest that the AMOC is weakening and has likely reached its slowest state in more than a millennium (22–31). However, the drivers of this decline remain debated: some studies attribute it primarily to natural variability (19, 32–35), whereas others suggest that the AMOC may be approaching a tipping point with the potential for an abrupt transition (36–38).

As the key pathway of AMOC's influence on regulating ocean heat transport (OHT) and ocean–atmosphere interactions, Atlantic surface currents contain the upper limb of the AMOC and provide an essential window for detecting AMOC variability (1, 3, 5, 17, 19, 39–42). Understanding systematic changes in these currents and their coupling with the AMOC is essential for clarifying how AMOC impact the surface ocean and for deciphering the evolution of the AMOC itself. However, the direct observation of AMOC remains too short to determine whether AMOC changes are accompanied by a systematic response of Atlantic surface circulation (3, 26, 43). Furthermore, surface circulation variability integrates multiple dynamical processes, making it difficult to isolate the specific signals driven by the AMOC (44).

Previous studies have largely relied on climate models to identify regions that are highly sensitive to AMOC variability as representation of the surface response, and construct AMOC proxy by sea surface temperature (SST), salinity (SSS) and height (SSH) (23, 43–48). While these proxies have offered valuable insights into the AMOC's surface footprint and long-term evolution beyond direct observations, they are highly heterogeneous across latitudes and variables (45, 49, 50). This lack of consistency leaves our understanding of basin-scale surface adjustments incomplete. Moreover, this model-based framework was fundamentally challenged in 2009, when the RAPID-MOCHA array recorded an abrupt 30% AMOC decline that far exceeded the interannual variability simulated in climate models used for IPCC assessment (22, 26) and most existing proxies failed to capture this event. Given that models cannot reproduce the full spectrum of real-world AMOC variability, the model-based identification may not be always robust under actual oceanic conditions, and relying on such predefined patterns risks overlooking the integrated dynamical features of the ocean surface. Consequently, there is a compelling need to move beyond model-based localized indicators and treat the Atlantic surface circulation as a unified, interconnected system.

Here, we represent the surface circulation as a linear superposition of a finite number of basin-scale spatiotemporal collective behaviors termed eigen microstates (EMs; see Methods) by applying Eigen Microstates Theory (EMT) (51–55) to directional angles of Atlantic surface currents—key descriptors of water-mass transport that are particularly sensitive to structural changes in ocean dynamics. EMs with finite eigenvalues capture spatiotemporal features of surface currents associated with different dynamical processes and can thus be interpreted as basin-scale dynamical phases of Atlantic surface circulation, enabling us to identify systematic reorganization and resolve how it is dynamically expressed. Within this framework, we identify a previously unrecognized phase of Atlantic surface circulation, which we term the Atlantic Convergence–Divergence Mode (ACDM). This phase is characterized by a convergence–divergence structure in the North Atlantic and coherent meridional flows in the South Atlantic. The temporal evolution of the ACDM reveals a pronounced step-like transition around 2009, indicating a major reorganization of the surface circulation system, marked by weakened vertical exchange and meridional transport accompanied by strengthened zonal flows. Notably, the interannual variability of the ACDM captures a sharp regime shift in 2009 that closely aligns with RAPID–MOCHA AMOC observations ($r = 0.69$, $p < 0.01$) (22, 26, 56, 57) and is strongly linked to AMOC-related ocean–atmosphere interactions. Building on this relationship, we develop an ACDM-based index that is highly sensitive to interannual AMOC variability and reveals abrupt transitions that are not captured by existing proxies (3, 23, 45, 47). Together, these results provide a concise and physically grounded framework for diagnosing basin-scale systematic changes in Atlantic surface currents and offer new insight into how AMOC variability propagates upward to reorganize the surface circulation, revealing that the recent AMOC decline is not a gradual trend but a non-linear, step-like transition that challenges our current understanding of its stability.

**Results**

**The Atlantic convergence-divergence mode**

The leading eigen microstate (EM1; Fig. S1A) accounts for the largest probability (68.4%) and manifests as a clockwise gyre in the North Atlantic and a counterclockwise gyre in the South Atlantic, consistent with the climatological mean circulation (58). Its temporal evolution (Fig. S1B) exhibits only weak fluctuations relative to the mean state, indicating that EM1 remains dynamically stable over the analysis period. After excluding this relatively stable, climatology-like component, the second eigen microstate (EM2; probability: 3.6%) emerges as the dominant dynamical phase. EM2 accounts for 11.5% of the residual probability, exceeding that of the third eigen microstate by a factor of 1.6 and that of the North Atlantic Oscillation (NAO)–related EM4 by a factor of 2.5 (Fig. S2–3). This establishes EM2 as the primary modulator of Atlantic surface current variability beyond the mean circulation, despite its comparatively small contribution to total probability. Such disproportionate dynamical importance is common in climate systems; for example, the El Niño–Southern Oscillation accounts for only 1.3% of the total probability in global surface air temperature eigen microstates (53). The prominence of EM2 therefore identifies it as a critical dynamical phase governing basin-scale reorganization of Atlantic surface currents. Accordingly, we focus on the spatiotemporal characteristics of EM2.

EM2 exhibits a distinctive convergence–divergence structure in the North Atlantic, with a key region spanning 36.3°N–43.0°N and 58°W–44°W (see Methods in Supplementary Materials), accompanied by coherent meridional flows in the South Atlantic (Fig. 1A). This spatial pattern captures meridional transport and associated vertical water exchange. During boreal summer, warm tropical waters are advected northward and subducted upon reaching the North Atlantic coastline within the key region, while a northward surface flow simultaneously develops in the South Atlantic. In boreal winter, this pattern reverses, characterized by southward meridional flow in the South Atlantic and divergence in the North Atlantic. Owing to its distinctive structure and dynamics, we refer to EM2 as the Atlantic Convergence–Divergence Mode (ACDM).

The temporal evolution of the ACDM displays a pronounced seasonal cycle with a quasi-periodicity of approximately one year (Fig. 1B–D). The positive phase corresponds to northward meridional flow and convergence in the North Atlantic, whereas the negative phase represents the opposite configuration. Over the period 1980–2022, the positive phase typically occurs from May to September, while the negative phase persists from October through April of the following year (Fig. 1C), indicating a recurrent phase exchange on an annual timescale. Notably, this seasonal cycle is asymmetric in both amplitude and duration. To quantify this asymmetry and isolate interannual variability, we compute the annual mean of the EM2 temporal evolution, denoted as $\phi_{ACDM}$. The resulting $\phi_{ACDM}$ exhibits a statistically significant decreasing trend over the past four decades ($\tau = -0.25$, $p < 0.01$) and undergoes a pronounced regime shift in 2009 (Fig. 1B). This shift reflects an interannual-scale weakening of the ACDM, manifested by reduced vertical water exchange and weakened meridional transport.

To further characterize this weakening, we compare the seasonal cycle amplitudes between a pre-transition period (T1: 1980–2008) and a post-transition period (T2: 2011–2022). The comparison reveals a clear structural adjustment of the seasonal cycle: the positive phase weakens substantially, whereas the negative phase intensifies in the post-transition period (Fig. 1D). This change indicates reduced downwelling and northward transport during boreal summer, accompanied by enhanced upwelling and southward flow in winter, highlighting a systematic reorganization of seasonal surface circulation following the 2009 transition.

**Structural reorganization of the ACDM**

To further characterize changes in the ACDM across the transition, we compare its spatial structure before and after the transition by computing eigen microstates separately for the two periods. For the ACDM (EM2), the characteristic convergence–divergence structure in the North Atlantic and the meridional flow in the South Atlantic persist across the transition (Fig. S4; consistency tests are provided in the Supplementary Materials). However, the relative importance of these structures—quantified by dominant rates (see Methods in Supplementary Materials)—undergoes a marked reorganization (Fig. 2A, B).

During the pre-transition period, the convergence–divergence structure in the North Atlantic and the meridional flows in the South Atlantic dominate over the zonal flows in the Tropical Atlantic (Fig. 2A; North Atlantic: 15°N–65°N; Tropical Atlantic: 15°S–15°N; South Atlantic: 60°S–15°S). This indicates that vertical water exchange and meridional transport play the primary dynamical roles prior to the transition. In contrast, after the transition, the importance of the North Atlantic convergence–divergence structure and the South Atlantic meridional flows decreases, while the importance of zonal flows in the Tropical Atlantic increases and surpasses that of the South Atlantic meridional component. These changes are not confined to a single velocity component but reflect a basin-scale structural reorganization: both meridional and zonal components weaken in the North and South Atlantic, whereas the Tropical Atlantic exhibits a pronounced strengthening in both components (Fig. 2B).

Additional insight is obtained by directly differencing the ACDM spatial patterns between the pre- and post-transition periods. The differences exhibit strong longitudinal coherence (Fig. S5), allowing the changes to be represented as zonally coherent latitudinal structures (Fig. 2C, D). The zonal component shows a marked westward strengthening in the tropical Atlantic (Fig. 2C). The meridional component weakens across most latitudes between 30°S and 30°N, indicating a substantial reduction in the northward branch of the ACDM. North of 30°N, the meridional component shifts toward a strengthened northward flow, opposing the southward branch of the original convergence–divergence structure and further signaling a weakening of the North Atlantic convergence–divergence configuration (Fig. 2D).

Together, these changes indicate a systematic reorganization of the ACDM spatial structure, characterized by weakened vertical water exchange and meridional transport, accompanied by strengthened zonal flows in the Tropical Atlantic. Such evidence suggests that the observed regime shift and subsequent weakening transcend a mere numerical decline in the index; rather, they reflect a fundamental spatial reconfiguration of the surface circulation. To quantify the associated change in system stability, we further compute the eigen microstate entropy of the Atlantic surface currents (55) (see Methods in Supplementary Materials; Fig. S6). We find that the entropy increases by approximately 13% after the transition, indicating reduced resilience and a shift toward a more unstable surface circulation state.

**Wind forcing and vertical thermal reorganization drive ACDM transition**

We next investigate the physical drivers of the ACDM by examining surface wind forcing and the vertical structure of ocean temperature in the Atlantic. The seasonal cycle of the ACDM is primarily driven by seasonal variations in sea surface winds (SSW; Fig. S7), which generate Ekman transport and wind-driven geostrophic currents (3, 16, 59, 60). On interannual timescales, SSW anomalies also exert a strong influence on ACDM variability, particularly in the North Atlantic. Regression analysis reveals a robust relationship between ACDM variability and wind anomalies over the subpolar gyre near the Azores, coinciding with the core region of the North Atlantic Oscillation (NAO) (Fig. 3A). Consistent with this link, the interannual ACDM index ($\phi_{ACDM}$) is significantly correlated with the accumulated NAO index ($r = 0.50$, $p < 0.01$). Negative NAO phases are associated with anomalous easterly winds over the subpolar North Atlantic and westerly winds over the subtropics, inducing southward meridional currents in the subtropical North Atlantic and northward currents in the subpolar region (61). However, despite the rapid return of the NAO to positive values following the extreme negative phase in 2009–2010 (3), $\phi_{ACDM}$ remained persistently negative (Fig. 3D). This persistence indicates that surface wind forcing alone cannot explain the ACDM transition, although it may have acted as a trigger for the shift (62).

We therefore examine the role of slower oceanic processes by analyzing the relationship between $\phi_{ACDM}$ and potential temperature across the vertical water column. Regression analysis without time lag reveals statistically significant correlations ($p < 0.05$) between $\phi_{ACDM}$ and potential temperature, with positive correlations at the surface (5 m) and negative correlations at depth (1479 m) within the key ACDM region (Fig. 3B,C). To quantify these signals, we compute spatially weighted annual mean potential temperatures and construct a surface temperature index and a deep temperature index for the region (Fig. 3E). The surface temperature index exhibits a pronounced decline beginning in 2004, reaching a minimum around 2010 before gradually recovering through 2017. In contrast, the deep temperature index begins to increase in 2005, peaks around 2009, and subsequently declines until 2019. These opposing trends are strongly anticorrelated ($r = -0.59$, $p < 0.01$, without time lag), indicating a weakening of the vertical temperature gradient that is consistent with enhanced upwelling in the key region during this period.

Extending the analysis across the full vertical range from 5 to 2174 m reveals a coherent vertical structure. The surface and upper-ocean layers (5–747 m) follow the surface cooling trend (Fig. S8A), whereas deeper layers (949–2174 m) exhibit warming similar to that observed at 1479 m depth (Fig. S8B). Together, these results indicate a basin-scale thermodynamic reorganization characterized by subsurface cooling and deep-ocean warming in the North Atlantic, consistent with a systematic weakening of vertical stratification. This restructuring of the vertical temperature profile likely modulated the stability of the surface circulation and played a key role in enabling the reorganization of the ACDM.

Taken together, our results indicate that the ACDM transition is not driven by a single forcing but instead emerges from the combined effects of slow oceanic adjustment and rapid atmospheric perturbations. The gradual evolution of the vertical temperature contrast establishes a low-frequency background state that preconditions the system, while episodic high-frequency anomalies in surface wind forcing—associated with the extreme negative NAO phase—act as rapid perturbations capable of triggering the transition. This behavior reflects a coupled ocean–atmosphere response, in which low-frequency subsurface thermal changes modulate the system's sensitivity to atmospheric variability, ultimately enabling a rapid reconfiguration of Atlantic surface circulation into a new quasi-stable state. Moreover, the coherent yet oppositely signed changes in surface and deep-ocean temperatures point to a direct link with the meridional overturning circulation, which governs vertical heat transport in the Atlantic.

**Relationship between the AMOC and the ACDM**

The coherent pattern of subsurface cooling and deep-ocean warming observed within the ACDM key region points to a close connection with the AMOC, which governs large-scale meridional heat redistribution in the Atlantic (3, 19, 32, 63). The subsurface cooling is consistent with the reported reversal of temperature trends around 2005, previously attributed to a decline in OHT driven by AMOC weakening (64). In addition, regression maps between $\phi_{ACDM}$ and potential temperature at 366 m and 459 m reveal a distinct dipole between the Gulf Stream region and the eastern subpolar gyre (Fig. S9), a pattern that closely resembles the canonical extratropical AMOC fingerprint (32, 48). A weakened AMOC reduces poleward OHT, leading to relative warming in the Gulf Stream region and cooling in the subpolar gyre. Through the wind–evaporation–SST feedback, this subpolar cooling can further propagate into the tropical North Atlantic (3, 19, 32, 48).

Beyond the upper ocean, AMOC weakening also diminishes the northward transport of North Atlantic Deep Water in the deep limb, limiting downward heat transfer from the upper ocean and resulting in heat accumulation near the interface between the upper and lower AMOC cells (32, 45, 57). Together, these processes provide a physically consistent explanation for the observed combination of subsurface cooling and deep-ocean warming in the ACDM key region. This vertical thermal reorganization, in turn, alters stratification and vertical exchange, thereby modulating

ACDM dynamics and highlighting the central role of the AMOC in shaping Atlantic surface circulation.

In addition to these thermodynamic linkages, the spatiotemporal characteristics of the ACDM show a strong correspondence with AMOC variability. The $\phi_{ACDM}$ is significantly correlated with direct AMOC observations from the RAPID–MOCHA array ($r = 0.69$, $p < 0.01$) and captures the abrupt AMOC decline in 2009 (26, 56). The identified ACDM key region is located near the western boundary, adjacent to the AMOC transition zone and the core region of deep Atlantic multidecadal variability (19, 65, 66). This region also overlaps with the descending branch of the AMOC's upper cell, spanning approximately 35°N–60°N, where lighter mode waters and thermocline waters are subducted (1, 15, 19, 67). The convergence–divergence structure of the ACDM is situated near the primary ingress and egress pathways of mode waters (1). This close spatial and dynamical alignment provides strong evidence for a tight coupling between the ACDM and the upper limb of the AMOC.

**A potential AMOC indicator defined by ACDM**

Given the strong dynamical linkage between the ACDM and the AMOC, we propose the interannual ACDM index ($\phi_{ACDM}$) as a potential indicator of AMOC variability. We evaluate the performance of $\phi_{ACDM}$ against direct AMOC observations and commonly used AMOC proxies (Fig. 4). Relative to the RAPID–MOCHA observations, $\phi_{ACDM}$ exhibits a strong and statistically significant correlation ($r = 0.69$, $p < 0.01$), comparable to the level of agreement often associated with fingerprints explicitly optimized to approximate AMOC variability. Importantly, beyond reproducing the overall variability, $\phi_{ACDM}$ uniquely captures the abrupt AMOC deceleration during 2009–2010. This correspondence is not the result of supervised statistical tuning or empirical curve fitting (3, 23, 36, 50, 68); rather, it emerges naturally from the collective behavior of Atlantic surface currents within a complexity science framework. This indicates that the ACDM distills intrinsic dynamical signatures of the overturning circulation, mapping the AMOC's step-like weakening as a self-organized outcome of surface circulation dynamics.

In contrast, existing AMOC proxies based on subpolar SST, salinity, or SSH (23, 44, 45) show substantially weaker correspondence with the RAPID–MOCHA record. At interannual timescales, these indices typically exhibit correlations below 0.3 without time lag and lack statistical significance (Table S1). Although they capture the broad multidecadal weakening of the AMOC, they fail to resolve the abrupt downturn in 2009 and the subsequent persistent reduction in strength (Fig. 4). Specifically, the subpolar SST and salinity indices reach their minima more than five years after the 2009 event, while the SSH-based index shows a delayed minimum in 2010 followed by a rapid rebound to above-normal values, inconsistent with the sustained weakening observed in direct measurements. These discrepancies underscore the superior sensitivity of $\phi_{ACDM}$ as a more sensitive and accurate proxy for detecting interannual-scale AMOC variability.

The enhanced performance of $\phi_{ACDM}$ likely arises from its grounding in a dynamically integrated variable that directly reflects surface current structure. In conventional fingerprints based on subpolar SST, salinity, or SSH over subpolar North Atlantic (23, 44, 45), the AMOC signal can be substantially masked by low-frequency internal variability associated with the Atlantic Multidecadal Oscillation (AMO), limiting their ability to resolve abrupt interannual shifts (3, 43). By incorporating the combined influences of wind stress, vertically distributed temperature anomalies, and vertical water exchange, $\phi_{ACDM}$ captures a broader spectrum of ocean–atmosphere interactions than single-variable proxies. In particular, it integrates high-frequency wind-driven forcing associated with the NAO and low-frequency thermohaline adjustments linked to AMOC variability. As such, the ACDM-based index provides a novel basin-scale framework for monitoring AMOC evolution from a complexity science perspective.

Our analysis confirms the long-term weakening trend of the AMOC but further demonstrates that this decline is not expressed as a smooth, gradual adjustment on decadal timescales. Instead, AMOC weakening is punctuated by step-like interannual transitions, with the 2009 regime shift marking a reorganization of the Atlantic circulation system. Consistent with previous studies, we also confirm the important role of episodic atmospheric forcing, such as extreme negative NAO events, in facilitating these transitions (62). Together, these results indicate that AMOC variability comprises both a gradual multidecadal component (23, 24) and abrupt interannual adjustments, the latter capable of driving basin-scale reorganization of Atlantic surface currents and pushing the system toward a more unstable state.

**Discussion and Conclusion**

Our analysis identifies a previously unrecognized dynamical phase of Atlantic surface circulation, which we term the Atlantic Convergence–Divergence Mode (ACDM), and demonstrates its strong alignment with observed variability of the Atlantic Meridional Overturning Circulation (AMOC). Consistent with direct RAPID–MOCHA observations, we find that the ACDM underwent a pronounced transition around 2009, characterized by a step-like decline in its temporal evolution and a systematic weakening of vertical exchange and meridional transport, accompanied by strengthened zonal flows in its spatial structure. This transition reflects a nonlinear adjustment of the coupled ocean–atmosphere system and provides a physically grounded explanation for the observed AMOC weakening. As the AMOC slows, the associated reduction in northward ocean heat transport alters meridional temperature gradients and surface wind forcing, driving a reorganization of the ACDM. Rather than representing a uniform attenuation of circulation strength, this response involves a redistribution of flow, with suppressed meridional exchange and enhanced zonal transport. Together, these findings demonstrate that Atlantic surface currents reorganize through tightly coupled cross-sphere interactions with the AMOC. The step-like nature of the transition, together with the observed 13% increase in eigen microstate entropy, suggests that the recent weakening of Atlantic surface circulation is not a simple linear decline but a step-like transition, in which the dominant EM is restructured while the probability distribution across EMs becomes more uniformed, indicating a more unstable state of the Atlantic surface currents.

The interannual variability of the ACDM robustly captures the 2009 regime shift (22, 26, 56), closely tracking RAPID–MOCHA AMOC observations ($r = 0.69$, $p < 0.01$) and reflecting AMOC-related ocean–atmosphere coupling mediated through the negative phase of the North Atlantic Oscillation and associated changes in ocean heat content. Importantly, this high-fidelity correspondence does not arise from empirical curve fitting but instead emerges from the collective dynamics of Atlantic surface currents, which encode intrinsic signatures of the overturning circulation. Building on this relationship, we propose an ACDM-based AMOC index that is more sensitive to interannual variability than existing regional proxies (3, 23, 44, 45). This index reveals that AMOC variability comprises not only gradual multidecadal changes but also abrupt interannual shifts capable of triggering basin-scale reorganization and system-wide weakening in the surface. These findings suggest that future simulations and assessments of AMOC stability should account for step-like transitions and the potential triggering role of extreme negative NAO events (62), rather than interpreting recent changes solely as smooth, gradual declines.

The AMOC-driven reorganization of surface currents has important implications for global heat redistribution and may influence large-scale climate phenomena such as the bipolar seesaw (69). Step-like declines on interannual timescales also raise the possibility of cascading impacts across the Earth system. Although our results do not indicate an imminent full-scale reversal or collapse of the AMOC, the observed sensitivity of the system to episodic atmospheric disturbances suggests that the AMOC may be more fragile and closer to a critical threshold than previously assumed (24, 36–38, 62). These findings underscore the urgency of improving monitoring of Atlantic circulation and assessing the potential consequences of continued weakening for climate stability, while also highlighting the need for proactive strategies to mitigate further irreversible changes in the ocean circulation system and to account for such step-like transitions in future assessments of Earth system stability.

## Materials and Methods

### Data sources

This study uses the following datasets: (1) Monthly mean zonal (u) and meridional (v) surface current velocities, together with monthly mean potential temperature fields, are obtained from the Global Ocean Data Assimilation System (GODAS) developed by the National Centers for Environmental Prediction (NCEP). The data are provided on a 0.33° × 1.0° (latitude × longitude) grid and span the period 1980–2022 (70); (2) Monthly mean zonal (u) and meridional (v) wind fields at 10 m height are taken from the NCEP/DOE AMIP-II Reanalysis (NCEP–DOE Reanalysis 2), produced by NCEP. These data have a horizontal resolution of 1.9° × 1.875° (latitude × longitude) and cover the period 1980–2022 (71); (3) Direct observations of the Atlantic Meridional Overturning Circulation are obtained from the RAPID–MOCHA array at 26°N (72), funded by the UK Natural Environment Research Council and freely available from www.rapid.ac.uk/rapidmoc; (4) The Atlantic Multidecadal Oscillation (AMO) index is defined as the annual mean of area-weighted sea surface temperature anomalies over the North Atlantic (73) (0°–70°N). The North Atlantic Oscillation (NAO) index is derived by projecting the NAO loading pattern onto daily anomalies of the 500-hPa geopotential height field over 0°–90°N. The loading pattern is defined as the leading mode of a rotated empirical orthogonal function (EOF) analysis applied to monthly mean 500-hPa height anomalies for the period 1950–2000 (74).

**Methods**

**Eigen Microstates Theory (EMT)**

The EMT offers a universal framework for the analysis of complex systems by extending Gibbs' ensemble theory in statistical physics (51–55). In a complex system of N agents, the collective states of all agents at time to constitute a microstate, while a statistical ensemble comprises all possible microstates in statistical physics. Utilizing the EMT, the statistical ensemble can be considered as the linear superposition of independent eigen microstates. For each eigen microstate whose probability is finite can be considered as one phase of this complex system.

The EMT has been effectively applied in various domains, such as predicting extreme El Niño/Southern Oscillation (ENSO) events (75), revealing how ENSO infuences SSS in the China Seas (76), evaluating state transitions and eutrophication in the Bohai Sea complex system (77), detecting phase transitions in turbulence as well as self-organized physical systems (78, 79). Its ability to reveal intrinsic structures and critical behaviors makes it an invaluable tool for exploring the underlying characteristics of diverse complex systems (80–84).

Here we consider the angles of Atlantic surface currents in the region of the Atlantic Ocean (Approximately 60°S–65°N, 98°W–20°E) as a complex system with N = 23437 grids. For each grid, the initial state combines the angle of the surface current velocity: $S_i(t) = [sin(\theta_i(t))\ cos(\theta_i(t))]^T$.

At a specific time t in the time period of M = 516, the microstate of the complex system can be symbolized by a vector $\boldsymbol{\delta S}(t)$:

$$\boldsymbol{\delta S}(t) = [S_1(t)\ S_2(t)\ \cdots\ S_N(t)]^T \\ = [sin(\theta_1(t))\ \cdots\ sin(\theta_N(t))\ cos(\theta_1(t))\ \cdots\ cos(\theta_N(t))]^T, \quad (1)$$

When all states are considered, the complex system can be represented as a statistical ensemble $A$, with its elements defined as $A_{it} = \frac{\delta S_i(t)}{C_0}$ and $C_0 = \sum_{t=1}^{M}\sum_{i=1}^{N} \delta S_i(t)^2$.

By employing singular value decomposition (SVD), the ensemble matrix can be factorized as follows:

$$\boldsymbol{A} = \boldsymbol{U\Sigma V^T}, \quad (2)$$

where $\boldsymbol{\Sigma}$ is a $2N \times M$ diagonal matrix with elemens:

$$\Sigma_{IJ} = \begin{cases} \sigma_I, I = J \leq r, \\ 0, \text{for others,} \end{cases} \quad (3)$$

where $r = min(2N, M)$, $\boldsymbol{U}$ and $\boldsymbol{V}$ are the obtained eigen microstates and their temporal evolution with time.

We can express the original microstates by eigen microstates as:

$$\delta \boldsymbol{S}(t) = \sqrt{C_0} \sum_{I=1}^{M} \sigma_I V_{tI} \boldsymbol{U}_I. \quad (4)$$

**The Dominant rate in Eigen Microstates**

It is essential to classify the importance of each component (meridional or zonal) in a specific spatial grid of the eigen microstate. Notably, each eigen microstate is an eigenvector, which means the $\sum_{i=1}^{2N} U_{Ii}^2 = 1$. Based on this fact, we can define the dominant rate of the component $i$ in eigen microstate $I$ as $\boldsymbol{D}_{Ii} = U_{Ii}^2$. The dominant rate stands for the percentage of the component $i$ in eigen microstate $I$, which indicates the importance of the zonal or meridional components in spatial grid $i$. For a specific spatial structure contain several spatial grids, we can sum the dominant rate of these grids. Employing the dominant rate in the eigen microstate, now we know the dominant structure in the eigen microstate.

**The Entropy of the Eigen Microstates**

In a complex system devoid of a dominant eigen microstate, the probabilities $p_I \to 0$ in the limit $N \to \infty, M \to \infty$. Conversely, when a non-vanishing $p_I = \sigma_I^2$ exists in the thermodynamic limit, the condensation of eigen microstate occurs, which is similar to Bose-Einstein condensation in a Bose gas (55). This condensation signifies a phase change, with the new phase characterized by the eigen microstate $U_1$. The order parameter for this phase transition is identified as $p_1$. From the probability distribution of eigen microstate, we can define eigen microstate entropy (55):

$$S_{EM} = - \sum_{I=1}^{N} p_I ln p_I. \tag{5}$$

$S_{EM}$ serves as an indicator of disorder by reflecting the statistical characteristics of the probability distribution of the eigen microstates. At ordered phase, system predominantly characterized by the largest eigen microstate $U_1$, for which the $p_1$ approaches unity. This minimizing $S_{EM}$ to nearly zero. In the disordered phase, the system lacks any dominant eigen microstates, resulting in a uniform probability distribution, i.e., $p_I \to 1/N$. Consequently, $S_{EM}$ reaches its maximum value with $ln\ N$. In general, a larger value of $S_{EM}$ signifies a higher degree of disorder in the system. The entropy, which first increases and then decreases, may indicate a potential critical transition between two ordered phases. And the derivative of the entropy may serve as the early warning signal of the phase transition.

In the angles of Atlantic surface currents complex system, we calculate the entropy of the eigen microstates for pre-transition and post-transition periods. To eliminate the influence of different time lengths across the ensemble, we normalized the entropy by the $ln(time\ length)$.

**The key region of the ACDM**

The key region of the ACDM is defined as the area which have convergence-divergence structure in the spatial pattern of ACDM and low strength (less than 0.45 of the absolute average value) in the divergence (Fig. S10), considering the uncertainty of the gulf stream, we choose the area for 36.3°N-43.0°N and 58°W-44°W.

**The interannual variation of the ACDM**

The interannual variation of the ACDM, $\phi_{ACDM}$ for each year is defined as the 12-month average of its temporal evolution from June to May of the following year. The annual window is defined from June to the following May, corresponding to the onset of the ACDM positive phase. Although this transition initiates in late spring, we adopt June as the reference point to ensure consistency with climatological seasonality.

**The annual average of climate variables**

To be aligned with $\phi_{ACDM}$, the annual average in this article is all defined as the 12-month average from June to May of the following year.

**Consistency Test of ACDM**

To verify that the ACDM derived from the entire period (January 1980–December 2022) can reliably represent the spatiotemporal changes captured by the part-time ACDM, we compared the full-time ACDM with the part-time ACDM. Since the part-time ACDM only characterizes the phase within a limited interval and cannot reflect the evolution over the whole record, we employed a projection approach: the spatial patterns of the part-time ACDM were projected onto the full-time ensemble to recover their long-term temporal evolutions:

$$A_{all}^{T} \cdot U_{ACDM}^{part} = \sigma_{ACDM}^{part} \cdot V_{ACDM-part}^{T}. \tag{6}$$

- $A_{all}^{T}$ is the ensemble matrix constructed from directional fields within the entire period,
- $U_{ACDM}^{part}$ represents the spatial patterns of ACDM derived from the specific period (e.g. the transition phase II),
- $\sigma_{ACDM}^{part}$ are the eigenvalues projected onto the part-time ACDM, and
- $V_{ACDM-part}^{T}$ describes the temporal evolution of the entire period associated with these projections by part-time ACDM.

For ACDM, we are particularly concerned with its interannual variation $\phi_{ACDM}$, so and the projected interannual variations from part-time ACDM are almost identical to those from the full-time ACDM (Fig. S11; all correlations > 0.98). Together, these results demonstrate that the full-time ACDM faithfully represents the spatiotemporal changes of the part-time ACDM, thereby justifying the use of the full-time ACDM evolution as a robust descriptor of ACDM variability.

**Mann-Kendall Test**

The Mann-Kendall test is a non-parametric test used to identify trends in time series data. It assesses whether there is a monotonic upward or downward trend over time without requiring the data to conform to any specific distribution (85).

**Lagged Pearson Correlation Coefficient**

The lagged Pearson correlation coefficient is an extension of the Pearson correlation coefficient, which measures the strength and direction of a linear relationship between two variables. For time series data, a lagged correlation assesses the relationship between one variable and another variable at different time lags. This is particularly useful in time series analysis when the effect of one variable on another may be delayed (86).

**The temperature index**

The temperature index is defined as the weighted average potential temperature in the key region detected by the ACDM of depth i m:

$$I_{im} = \overline{T_{im}^{Key\ Region}}, \tag{8}$$

Specifically, we define the $I_{5m}$ as the surface temeperature index and the $I_{1479m}$ as the deep temperature index.

**The AMOC proxies:**

The direct observation AMOC index is the annual average of the data measured by RAPID-MOCHA array. The subpolar gyre SST index (23), the subpolar upper ocean salinity index (45) and sea level proxy (44) are directly from the original article.

**Code availability**

The computer code used for analysis in this study was written in Python. Specific codes used in this study can be made available to interested readers upon request.

**Data availability**

The data that support the findings of this study are freely available. The GODAS dataset is available at https://psl.noaa.gov/data/gridded/data.godas.html. The NCEP/DOE Reanalysis II data is available at https://psl.noaa.gov/data/gridded/data.ncep.reanalysis2.html. The ORAS5 dataset is available at https://cds.climate.copernicus.eu/portfolio/dataset/reanalysis-oras5. The RAPID-MOCHA Array data is available at www.rapid.ac.uk/rapidmoc. The AMO index is available at https://psl.noaa.gov/data/timeseries/AMO/. The NAO index is available at https://www.cpc.ncep.noaa.gov/products/precip/CWlink/pna/norm.nao.monthly.b5001.current.ascii.table. The shapefile of the Atlantic Ocean is available at https://www.marineregions.org/gazetteer.php?p=details&id=1902.

**Fundings:**

National Key R&D Program of China grant 2023YFE0109000

National Key R&D Program of China grant 2025YFF0517304

National Natural Science Foundation of China grant 12135003, grant T2525011, grant 42450183, grant12275020, grant 42461144209

European Union's Horizon Europe research and innovation program grant No. 101137601

Fundamental Research Funds for the Central Universities

Deutsche Forschungsgemeinschaft (DFG): Climate Resilience under Zero-Emission Commitment Scenarios, BO 4455/2-1 (eBer-24-62950; RS-2024-00438471)

Volkswagen Foundation

Horizon Europe research and innovation programme under grant agreement No.'s 101137601 (ClimTip) and 101184070 (Past2Future).

Advanced Research and Invention Agency (ARIA) – AdvanTip (SCOP-PR01-P003)

Tsinghua University (100008001)

**Acknowledgements:**

We acknowledge current data derived from the Global Ocean Data Assimilation System (GODAS) developed at National Centers for Environmental Prediction (NCEP) and the data from the RAPID AMOC monitoring project is funded by the Natural Environment Research Council and U.S. National Science Foundation. We also thank Yu Sun, Xuan Ma, Shang Wang, Sheng Fang and Xu Wang for their assistance in discussing and/or analyses.

**Author Contributions:**

H.H. and X.C. contributed to the original idea. H.H and X.C developed the theoretical framework and drafted the figures and writing of this paper. All authors discussed and interpreted the results and provided input to the manuscript.

**Competing interests**

The authors declare that they have no competing interests.

# Figures

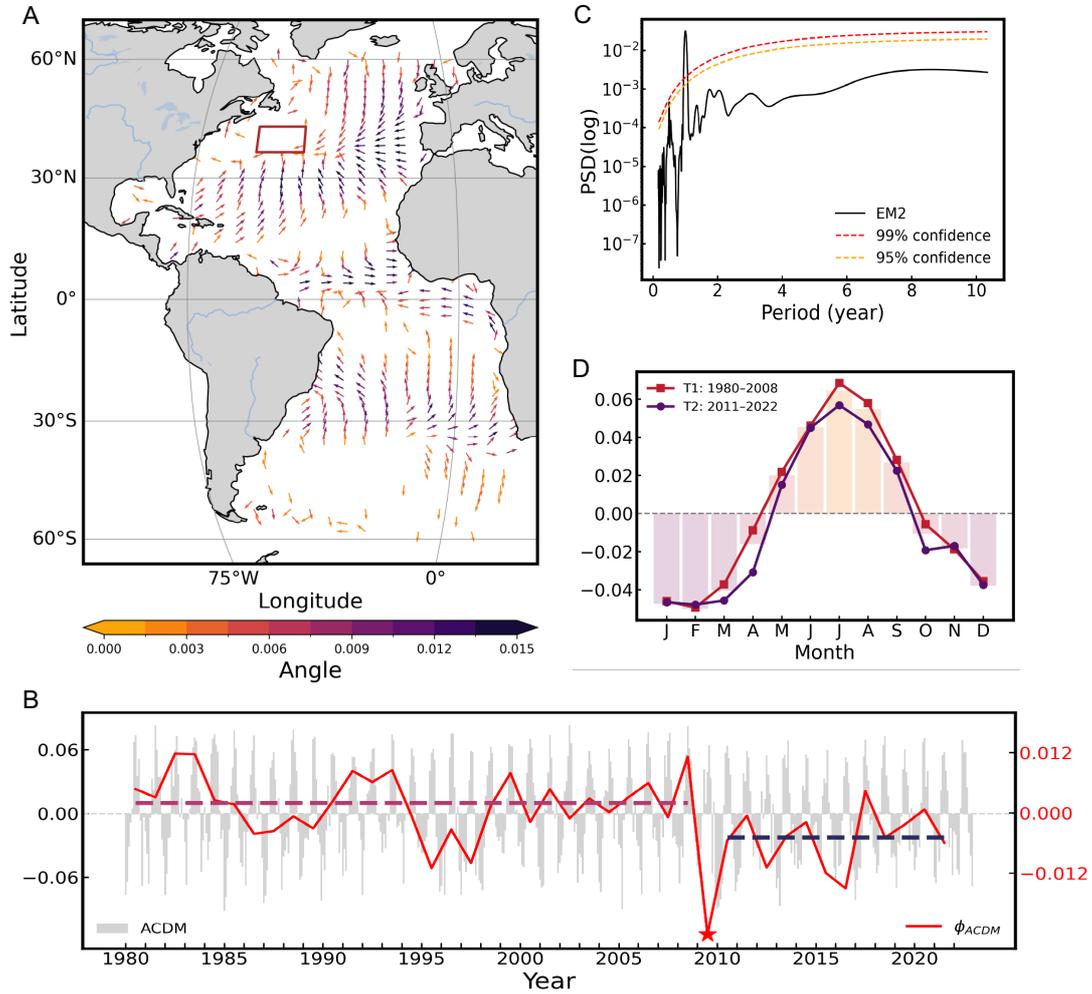

**Fig. 1. Spatiotemporal characteristics, spectral properties, and seasonal cycle of the Atlantic Convergence–Divergence Mode (ACDM).** The ACDM of Atlantic surface currents derived from GODAS for the period 1980–2022. (**A**) Spatial pattern of the ACDM. Arrows indicate flow direction, and shading denotes magnitude. The brown box marks the key region. Vectors with magnitudes smaller than 0.002 are omitted for clarity. (**B**) Temporal evolution of the ACDM. Gray bars show the EM2 time series, and the red curve represents its interannual component ($\phi_{ACDM}$). The dotted purple line denotes the mean value of $\phi_{ACDM}$ before the 2009 transition, and the dotted dark blue line denotes the mean after the transition. (**C**) Power spectral density of the ACDM. The red and orange curves indicate the 99% and 95% confidence levels, respectively, relative to a red-noise spectrum. (**D**) Seasonal cycle of the ACDM. Bars show the climatological seasonal cycle for the full period (1980–2022), the red curve corresponds to the pre-transition period (T1: 1980–2008), and the purple curve corresponds to the post-transition period (T2: 2011–2022).

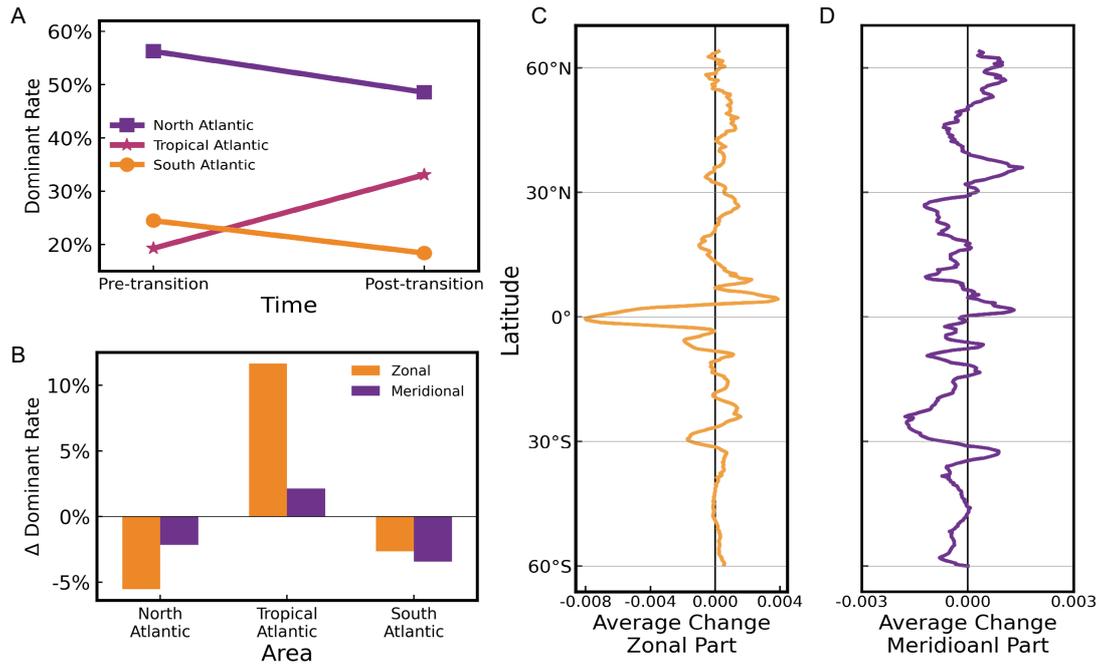

**Fig. 2. Structural reorganization of the Atlantic Convergence–Divergence Mode (ACDM).** (**A**) Dominant rates of the ACDM across different Atlantic sectors during the pre-transition and post-transition periods. (**B**) Differences in dominant rates between the two periods, separated into meridional and zonal components across Atlantic sectors. (C, D) Zonal-mean differences between pre-transition and post-transition ACDM spatial patterns: zonal component (**C**) and meridional component (**D**).

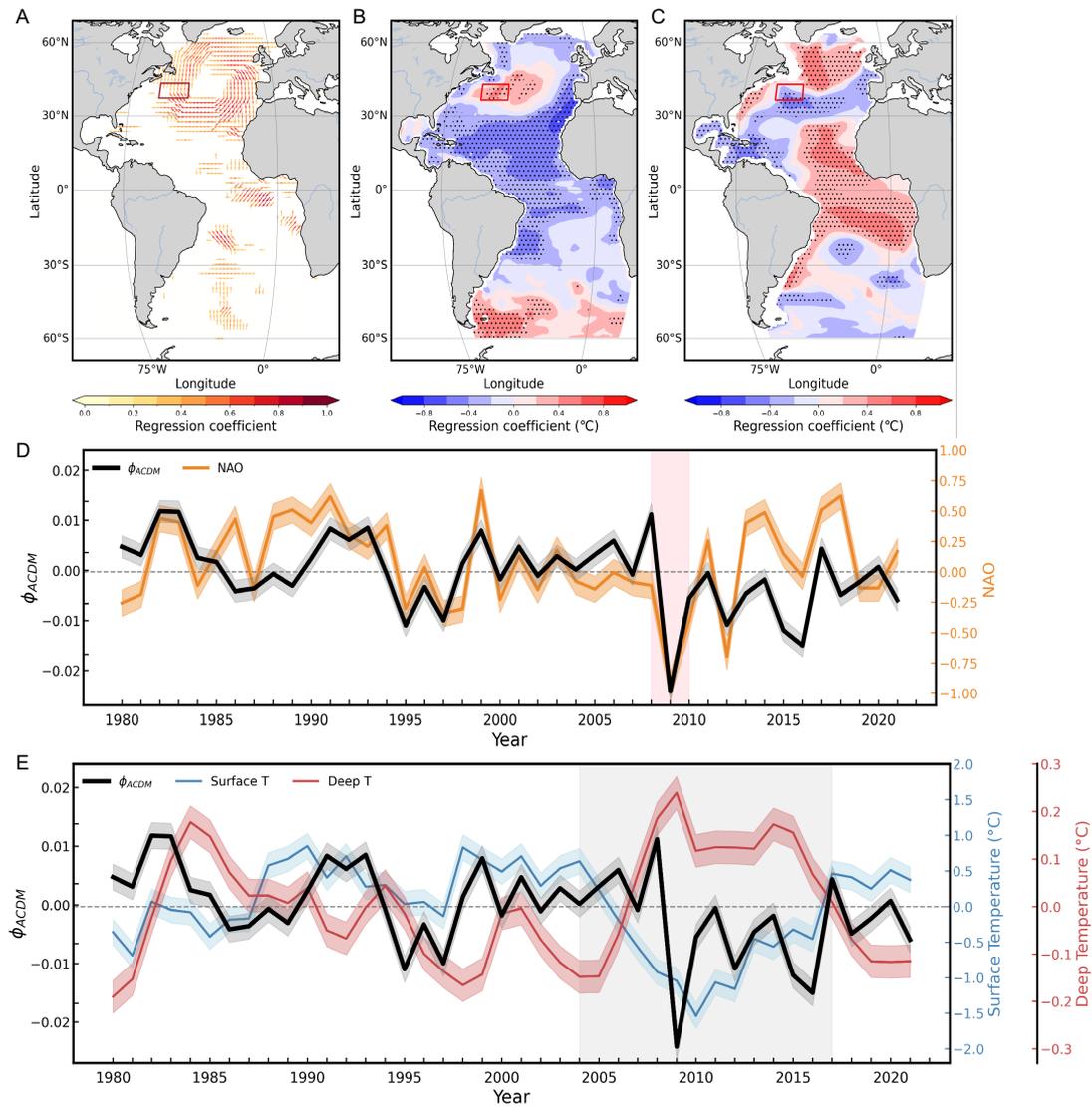

**Fig. 3. Interannual drivers of the ACDM index $\phi_{ACDM}$.** (**A**) Regression of sea surface wind anomalies onto $\phi_{ACDM}$. Arrows indicate wind direction, and color denotes the magnitude of the correlation. All vectors are statistically significant ($p < 0.05$). The brown box marks the key ACDM region. (**B**) Regression of potential temperature at 5 m depth onto $\phi_{ACDM}$, showing a positive signal within the key region. Stippling indicates statistical significance at the 95% confidence level ($p < 0.05$) based on a two-tailed Student's t-test. (**C**) Same as (**B**), but for potential temperature at 1479 m depth, showing a negative signal within the key region. (**D**) Comparison between $\phi_{ACDM}$ (black curve) and the annual NAO index (orange curve), calculated by averaging the NAO index from June to the following May. The pink shading highlights the period 2004–2017. (**E**) Comparison between $\phi_{ACDM}$ (black curve), the surface temperature index (blue curve), and the deep temperature index (red curve), derived from spatially weighted annual mean potential temperatures within the key region at 5 m and 1479 m depth, respectively. The light gray shading highlights the period 2004–2017.

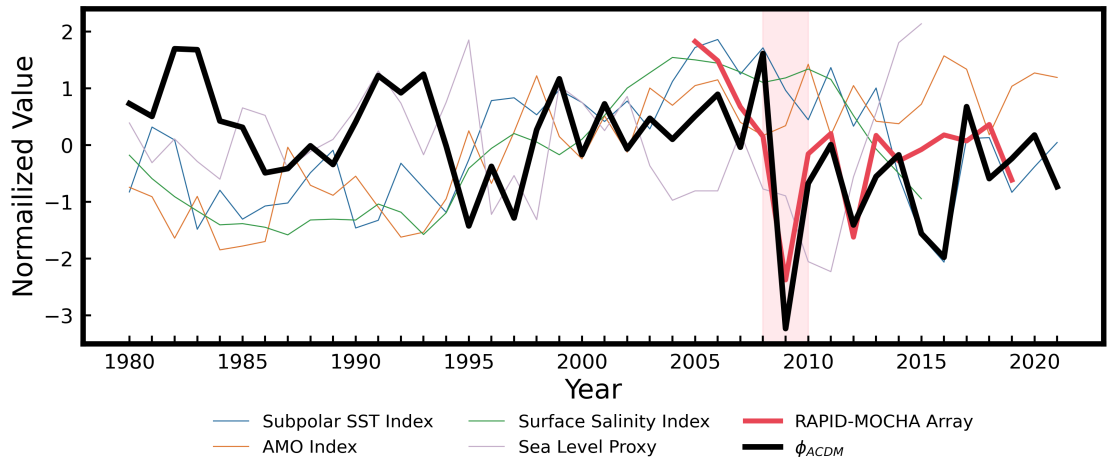

**Fig. 4. Performance of the ACDM-based index $\phi_{ACDM}$ compared with existing AMOC proxies.** Comparisons of multiple AMOC indices with direct AMOC observations from the RAPID-MOCHA array at 26°N. The thick black curve denotes $\phi_{ACDM}$, and the red curve shows the RAPID-measured AMOC strength. The thin blue, green, purple, and orange curves represent the subpolar gyre SST index, surface salinity index, sea level–based proxy, and annual AMO index, respectively. Only $\phi_{ACDM}$ captures the abrupt AMOC weakening observed in 2009. The pink shading highlights the period associated with the ~30% reduction in AMOC strength measured by the RAPID–MOCHA array. The correlation coefficient between $\phi_{ACDM}$ and the RAPID–MOCHA AMOC record is 0.69.